# Robust and High-Fidelity Controlled Two-Qubit Gates via Asymmetric Parallel Resonant Excitation

LICHENG LIN [1,2,†], JIZE HAN [3,†], PENG ZHU [1,2], ZIYU WANG [1,2], YING YAN [1,2,*], JIE LU [4,5,6], AND ZHIGUO HUANG [3,7]

1. School of Optoelectronc Science and Engineering & Collaborative Innovation Center of Suzhou Nano Science and Technology, Soochow University, Suzhou 215006, China;
2. Key Lab of Advanced Optical Manufacturing Technologies of Jiangsu Province & Jiangsu Key Laboratory of Flexible Optoelectronics and Micro-Nano Manufacturing & Key Lab of Modern Optical Technology of Education Ministry of China, Soochow University, Suzhou 215006, China;
3. China Mobile (Suzhou) Software Technology Co., Ltd., Suzhou 215163, China;
4. Department of Physics, Shanghai University, 200444 Shanghai, China;
5. Institute for Quantum Science and Technology, Shanghai University, 200444 Shanghai, China;
\* *yingyan@suda.edu.cn*
6. *lujie@shu.edu.cn*
7. *huangzhiguo15@mails.ucas.ac.cn*
†These authors contributed equally.

**Abstract:** Implementing high-fidelity controlled two-qubit gates in dipole-dipole interacting systems, such as rare-earth-ion crystals, in hindered by spectral inhomogeneity and weak coupling. Existing method often rely on detuned pulses, making them susceptible to frequency errors and AC Stark shifts. We propose a robust resonant scheme for arbitrary controlled two-qubit gates that utilizes asymmetric excitation and pulse engineering to achieve decoupled, parallel qubit control. Simulations on rare-earth-ion ensemble qubits demonstrate gate fidelities exceeding 99% within a ±170 kHz detuning range with off-resonant excitation below 0.2%. This approach offers a robust, scalable route for quantum computing in spectrally crowded systems.

## 1. INTRODUCTION

Quantum computing, enabled by superposition and entanglement, is emerging as a revolutionary paradigm for solving problems intractable to classical computers, such as integer factorization and unstructured database search [1,2]. A key requirement for harnessing its computational potential is the realization of high-fidelity quantum gates. Since any quantum algorithm can be decomposed into single- and two-qubit operations, the performance of these elementary gates directly determines the reliability and scalability of quantum computation.

Significant advances have been achieved in the realization of high-fidelity single-qubit gates across various physical platforms, including superconducting circuits, trapped ions, and neutral atoms. This progress has been driven by a range of sophisticated control techniques, most notably non-adiabatic geometric quantum computation (NGQC), non-adiabatic holonomic quantum computation (NHQC), and the capture-and-release method [3-8]. These approaches have been systematically optimized and experimentally validated [9-12], delivering fast, robust, and high-fidelity single-qubit control that forms a solid foundation for scalable quantum information processing.

In contrast to the relatively mature and platform-agnostic nature of single-qubit control, the implementation of two-qubit gates relies heavily on platform-specific physical interactions. For instance, superconducting circuits employ tunable capacitive coupling enabled by adjustable electric fields to construct multi-qubit gates [13]. Trapped-ion systems utilize shared vibrational modes to mediate interactions, giving rise to gates such as the Mølmer–Sørensen and the light-shift gates [14,15]. In neutral atom platforms like Rydberg atoms, strong dipole–dipole interactions between Rydberg states [16] enable the blockade effect [17,18], which has motivated the development of both time-sequenced (serial) [19] and non-time-sequenced (parallel) entangling schemes [20-25]. Although more intricate in design, parallel schemes generally offer superior speed and efficiency, with examples including Rydberg anti-blockade combined with asymmetric excitation [24] and complex hyperbolic-secant pulses for arbitrary controlled gates [25].

A common challenge across many advanced two-qubit schemes is their reliance on detuned pulses or effective off-resonance conditions. These approaches demand precise frequency control and are susceptible to errors such as AC Stark shifts [25]. Here, we overcome this limitation by introducing a resonant scheme for arbitrary controlled two-qubit gates based on asymmetric excitation. While the scheme is conceptually applicable to various platforms with dipole-dipole interactions, we demonstrate its effectiveness here using the ensemble rare-earth-ion (REI) platform—specifically, $Eu^{3+}$ ions doped in a $Y_2SiO_5$ crystal.

REI doped crystals constitute a compelling quantum platform due to their exceptionally long coherence times [26], outstanding spectral stability, and strong potential for scalable integration, making them attractive for quantum memory [27] and gate-based quantum processing [26,28]. High-fidelity single-qubit gates based on NGQC and NHQC protocols have already been proposed [29-31] and demonstrated in these systems [32], confirming their compatibility with advanced control techniques. Furthermore, both time-sequenced [33,34] and detuned non-time-sequenced entangling schemes [25] have been developed for REI platform, underscoring its potential for scalable multi-qubit operations.

Nevertheless, achieving high-fidelity two-qubit gates in ensemble REI system remains challenging. For example, in the case of the ensemble qubit in a $Eu^{3+}:Y_2SiO_5$ crystal. The intrinsic spectral inhomogeneity of an ensemble, for example, a detuning range of about 340 kHz in $Eu^{3+}:Y_2SiO_5$ [26], requires gate operations to be highly robust against frequency variations. Simultaneously, the relatively weak dipole–dipole interaction strength, together with a spectrally dense environment, imposes stringent constraints on pulse design: the pulses must deliver a sufficient Rabi frequency to ensure gate efficiency, while maintaining off-resonant excitation below 5% for transitions detuned by about 8.9 MHz from the target frequency [35].

To address these challenges, we develop a non-time-sequenced arbitrary two-qubit gate protocol that combines resonant asymmetric excitation, tailored pulse-envelope engineering, and a phase-compensation procedure. Through numerical optimization of the pulse shapes, we simultaneously meet the dual requirements of strong robustness against frequency detuning and low off-resonant excitation. This work fills the gap in the resonance controlled two-qubit gate scheme in this system, highlighting the potential of rare-earth ions as a versatile platform for scalable quantum computing.

The article is structured as follows. Section 2 introduces the theoretical model and pulse ansatz. Section 3 presents numerical simulation results and a comprehensive analysis of gate performance. Finally, Section 4 summarizes the key findings and discusses broader implications and future directions.

## 2. THEORETICAL MODEL

We consider a two-qubit system in which each qubit is modeled as a three-level system. The control qubit is driven by a single external field $\Omega_c(t)$ targeting on the transition of $|0\rangle$-$|e\rangle$, in the mean while the target qubit is simultaneously driven by two optical fields, $\Omega_{0t}(t)$ and $\Omega_{1t}(t)$, as illustrated in Fig. 1. The two qubits interact with each other via the permanent dipole–dipole coupling. When population of one qubit is changed between the ground and excited states, the resulting change in its permanent dipole moment induces a shift in the resonance frequency of the other qubit, thereby preventing its simultaneous excitation. This phenomenon is known as the dipole blockade effect, whose magnitude is characterized by the interaction strength $V$.

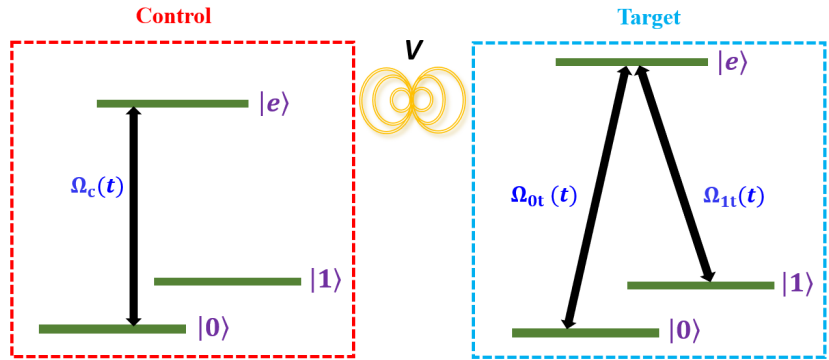

**Fig. 1.** Schematic energy-level diagram of the three-level models for a two-qubit system. Each qubit consists of two ground states $|0\rangle$ and $|1\rangle$, which are coupled indirectly via a common external state $|e\rangle$. The control qubit is addressed only on the $|0\rangle \leftrightarrow |e\rangle$ transition by the external field $\Omega_c(t)$, while the target qubit is addressed on both ground-state transitions, $|0\rangle \leftrightarrow |e\rangle$ and $|1\rangle \leftrightarrow |e\rangle$, by the fields $\Omega_{0t}(t)$ and $\Omega_{1t}(t)$, respectively. The two qubits interact with each other via a permanent dipole–dipole interaction $V$.

The two external fields $\Omega_{0t}(t)$ and $\Omega_{1t}(t)$ for the target qubit, are defined as $\Omega_{0t}(t) = \Omega_t(t)\sin(\theta/2)e^{i\varphi_{0t}}$ and $\Omega_{1t}(t) = -\Omega_t(t)\cos(\theta/2)e^{i\varphi_{1t}}$, where $\Omega_t(t)$ denotes the effective Rabi frequency envelope of the two driving field applied to the target qubit and it satisfies $\Omega_t(t) = \sqrt{\Omega_{0t}(t)^2 + \Omega_{1t}(t)^2}$. $\theta$ and $\varphi_{0t,1t}$ are time-independent angles in the range of $[0, 2\pi]$, describing the relative amplitude and phase components of the time-dependent Rabi frequencies, respectively. Based on these parameters, the individual Hamiltonian of the two qubits, $H_c$ and $H_t$, and the interaction Hamiltonian between them, $H_v$, reads:

$$H_c = \Omega_c(t)|0\rangle\langle e| + \text{H.c.}, \tag{1}$$

$$H_t = \Omega_{0t}(t)|0\rangle\langle e| + \Omega_{1t}(t)|1\rangle\langle e| + \text{H.c.}, \tag{2}$$

$$\text{and } H_v = V|ee\rangle\langle ee|, \tag{3}$$

respectively, where H.c. represents Hermitian Conjugate. To simplify the calculation of the evolution operator, we change the basis of the target qubit in Eq. (2) from $\{|0\rangle, |1\rangle, |e\rangle\}$ to $\{|b\rangle, |d\rangle, |e\rangle\}$, where $|b\rangle = \sin(\theta/2)|0\rangle - \cos(\theta/2)e^{i\phi}|1\rangle$ and $|d\rangle = \cos(\theta/2)e^{-i\phi}|0\rangle + \sin(\theta/2)|1\rangle$ are called the bright and dark states, respectively, with $\phi = \varphi_{1t} - \varphi_{0t}$ being the relative phase between $\Omega_{0t}(t)$ and $\Omega_{1t}(t)$. As a result, the $H_t$ in Eq. (2) turns to a simple form of $H_t = \Omega_t(t)|b\rangle\langle e| + \text{H.c.}$.

Now the Hamiltonian of the whole system can be constructed as:

$$H(t) = H_c \otimes I_t + I_c \otimes H_t + H_v, \tag{4}$$

where $\otimes$ represents the tensor product, and $I_{c,t}$ an identity operator of the control or target qubit.

We then transition to the rotating frame via the unitary transformation [36] $\mathcal{U} = \exp(-iH_v t)$. Given that the dipole–dipole interaction strength $V$ is much larger than the Rabi frequencies involved, we can apply the rotating-wave approximation (RWA) and eliminate the term corresponding to the doubly excited state $|ee\rangle$, yielding the simplified Hamiltonian:

$$\begin{aligned} H(t) = &\Omega_c(t)(|0b\rangle\langle eb| + |0d\rangle\langle ed|) \\ &+ \Omega_t(t)e^{i\varphi_{0t}}(|0b\rangle\langle 0e| + |1b\rangle\langle 1e|) + \text{H.c.}. \end{aligned} \tag{5}$$

To enable parallel manipulation of the control and target qubits and engineer the evolution path, we introduce an asymmetric excitation scheme where $\Omega_c(t)$ is proportional to $\Omega_t(t)$ as $\Omega_c(t) = \sqrt{3}\Omega_t(t)e^{i\varphi_c}$. $\varphi_c \in [0, 2\pi]$ represents the phase of $\Omega_c(t)$. Under this configuration, the Hamiltonian can be rewritten as:

$$\begin{aligned} H(t) = &2\Omega_t(t)|0b\rangle\langle T|) \\ &+ \sqrt{3}\Omega_t e^{i\varphi_c}|0d\rangle\langle ed|) + \Omega_t(t)|1b\rangle\langle 1e| + \text{H.c.}, \end{aligned} \tag{6}$$

where $|T\rangle = (\sqrt{3}|eb\rangle e^{i\varphi_c} + |0e\rangle e^{i\varphi_{0t}})/2$ represents an intermediate state that emerges during the evolution.

*A. Arbitrary Controlled Two-Qubit Gates*

In this section, we demonstrate the construction of universal arbitrary controlled two-qubit gates based on the Hamiltonian in Eq. (6). Inspired by the orange-slice scheme in the single-qubit gates[37], we aim for implementing the singe qubit gate on the target qubit by utilizing the orange-slice scheme on condition of the control qubit being state $|1\rangle$, which is a bit complicated in the non-sequential scenario. To achieve this, we divide the evolution into two sequential time segments. In each segment, the pulse area remains constant, satisfying

$$\int_{t_0}^{t_0+\tau} \Omega(t)dt = \pi/2, \tag{7}$$

where $t_0$ represents the starting time in each segment, and the phases are configured as follows:

$$\text{For } t \in [0, \tau], \quad \begin{cases} \varphi_c = 0, \\ \varphi_{0t} = 0, \varphi_{1t} = \phi; \end{cases} \tag{8}$$

$$\text{For } t \in [\tau, 2\tau], \quad \begin{cases} \varphi_c = \pi, \\ \varphi_{0t} = \pi + \gamma, \varphi_{1t} = \phi + \pi + \gamma, \end{cases} \tag{9}$$

where $\tau$ denotes the duration of each pulse, and $\gamma$ represents the geometric phase accumulated from the phase contributions of both segments. Based on the Hamiltonian in Eq. (6) and driving Rabi frequency in Eqs. (7)-(9), the time evolution operator of the system is $U(t, 0) = \hat{P} e^{-i \int_0^t H(t')dt'} = e^{-i \int_0^t H(t')dt'}$, where the time ordering operator $\hat{P}$ is omitted since the Hamiltonian in Eq. (6) at any two distinct times commute with each other. Specifically, the time evolution operator for each segment is

$$\begin{aligned} U(\tau, 0) = &-|0b\rangle\langle 0b| - |T\rangle\langle T| - i(|1b\rangle\langle 1e| + |1e\rangle\langle 1b|) \\ &+ \cos\left(\frac{\sqrt{3}\pi}{2}\right)(|0d\rangle\langle 0d| + |ed\rangle\langle ed|) \\ &- i\sin\left(\frac{\sqrt{3}\pi}{2}\right)(|0d\rangle\langle ed| + |ed\rangle\langle 0d|) \\ &+ |1d\rangle\langle 1d| + |ee\rangle\langle ee|, \end{aligned} \tag{10}$$

and

$$\begin{aligned} U(2\tau, \tau) = &-|0b\rangle\langle 0b| - |T\rangle\langle T| + i(e^{i\gamma}|1b\rangle\langle 1e| + e^{-i\gamma}|1e\rangle\langle 1b|) \\ &+ \cos\left(\frac{\sqrt{3}\pi}{2}\right)(|0d\rangle\langle 0d| + |ed\rangle\langle ed|) \\ &+ i\sin\left(\frac{\sqrt{3}\pi}{2}\right)(|0d\rangle\langle ed| + |ed\rangle\langle 0d|) \\ &+ |1d\rangle\langle 1d| + |ee\rangle\langle ee|. \end{aligned} \tag{11}$$

Then the final time-evolution operator, with $|1d\rangle$ and $|ee\rangle$ decoupled from the system, reads:

$$\begin{aligned} U(2\tau,0) &= U(2\tau,\tau)U(\tau,0) \\ &= e^{i\gamma}|1b\rangle\langle 1b| + e^{-i\gamma}|1e\rangle\langle 1e| \\ &\quad + e^{i2\pi}|0d\rangle\langle 0d| + e^{i2\pi}|ed\rangle\langle ed| + e^{i2\pi}|0b\rangle\langle 0b| \\ &\quad + |1d\rangle\langle 1d| + |ee\rangle\langle ee| + 3|eb\rangle\langle eb|/4 + e^{i\gamma}|0e\rangle\langle 0e|/4. \end{aligned} \quad (12)$$

To illustrate the controlled gates, we would project the $U(2\tau,0)$ in Eq. (12) onto the computational basis $\{|00\rangle,|01\rangle,|10\rangle,|11\rangle\}$, which gives:

$$U(2\tau,0) = |0\rangle\langle 0| \otimes I + |1\rangle\langle 1| \otimes e^{-i\frac{\gamma}{2}} e^{i\frac{\gamma}{2}\hat{n}\cdot\vec{\sigma}}, \quad (13)$$

where $\hat{n} = (sin\theta cos\phi, sin\theta sin\phi, cos\theta)$ denotes a unit vector, and $\vec{\sigma} = (\sigma_x, \sigma_y, \sigma_z)$ the Pauli matrices of the target qubit. Eq. (13) indicates that on condition of the state, $|1\rangle$ or $|0\rangle$, of the control qubit, the target qubit either executes a rotation around the axis $\hat{n}$ by an angle of $\gamma$ up to a global phase factor $e^{-i\frac{\gamma}{2}}$, or remains unchanged. In this way one can realize arbitrary controlled two-qubit gates with appropriate values of key parameters. Examples of several commonly used controlled two-qubit gates are listed in Table 1.

Table 1.  Parameters for some controlled two-qubit quantum gates. CNOT: Controlled-Not gate. CH: Controlled-Hadamard gate. CZ, CS, CT: Controlled-Phase gate of three phases

| Gates | $\gamma$ | $\phi$ | $\theta$ |
|---|---|---|---|
| CNOT | $\pi$ | 0 | $\pi/2$ |
| CH | $\pi$ | 0 | $\pi/4$ |
| CZ | $\pi$ | 0 | 0 |
| CS | $\pi/2$ | 0 | 0 |
| CT | $\pi/4$ | 0 | 0 |

To provide an intuitive understanding of the gate dynamics, we illustrate the effective evolution paths on Bloch spheres in Fig. 2. When the control qubit is initially in state $|1\rangle$, it is effectively decoupled from the external field, allowing the target qubit to evolve independently. During the first interval $[0,\tau]$, the target state evolves upward along the orange trajectory, while during the second interval $[\tau,2\tau]$, it descends along the red trajectory. Together, these segments form an orange-slice-shaped path[37], resulting in the accumulation of a geometric phase $\gamma$ on the target qubit.

In contrast, when the control qubit is in state $|0\rangle$, the two qubits undergo entangled evolution, which cannot be fully depicted on the Bloch sphere. Nevertheless, as described by Eqs. (10) and (11), the evolution within the computational subspace accumulates a $\pi$ phase in each segment, giving a total enclosed phase of $2\pi$. Consequently, both the control and target qubits return to their initial states by the end of the sequence.

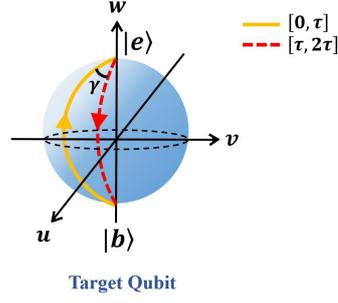

**Fig. 2.** Evolution paths of the target qubit on the Bloch sphere. On condition of the control qubit being in the $|1\rangle$ state, the target state evolves along an "orange-slice" trace and accumulates a geometric phase $\gamma$.

In the following passage, we will apply the controlled gates scheme to the ensemble REI system to illustrate the gate performance. In theory, any pulses satisfying Eq. (8) and (9) are able to execute the gate operations. However, in practice the gate performance, such as the robustness against the systematic errors and unwanted disturbance from the off-resonant excitations, is highly affected by the exact dynamics of the evolution. Therefore, it is necessary and critical to carefully engineer the waveforms of the pulses.

## B. Ansatz of Pulses

We propose the effective Rabi frequency envelope $\Omega_t(t)$ takes the form of parametric cosine harmonics (PCH), specifically:

$$\Omega_t(t) = \frac{\pi}{2\tau} + \sum_{k=1}^{s} a_k \frac{k\pi}{\tau} \cos\left(\frac{k\pi t}{\tau}\right), t \in (0, 2\tau), \tag{14}$$

where $a_k$ denotes the weights assigned to the $k_{th}$ harmonic and $s$ the maximum value of $k$. This pulse naturally meets the pulse area condition shown in Eq. (7), ensuring that any choice of the coefficients $a_k$ can realize the desired controlled gate. However, the choice of $a_k$ has a strong impact on the gate's performance, thereby offering a useful degree of freedom for performance optimization. Similar waveforms have shown remarkable performance in single-qubit gate operations [30,31]. However, extending them to simultaneously driven controlled two-qubit gates is more challenging due to the enlarged Hilbert space and more complicated dynamical evolution.

Practically, a zero starting and ending point of the Rabi frequency in each segment is preferable, as any steep rising or falling edge in time could bring abundant frequency components, which would potentially induce unwanted excitations. This condition imposes boundary conditions to $a_k$ as follows:

$$a_1 + 3a_3 + 5a_5 + \cdots + (2q-1)a_{2q-1} = 0, \tag{15}$$

$$2a_2 + 4a_4 + 6a_6 + \cdots + (2q)a_{2q} = -0.5, \tag{16}$$

where $q = 1, 2, 3, \ldots$ In principle $q$ is unbounded. However, higher values would increase the complexity of the pulses and potentially lead to oscillations of the state evolution yet with diminishing gains in fidelity. Here we set $q = 2$ in each segment to achieve a reasonable tradeoff between efficiency and simplicity.

## C. Compensation Pulses

The theoretical scheme presented above assumes that the external light to be perfectly resonant with the atomic transition. However, frequency detuning or drift is often unavoidable in real experiments. Under such imperfect conditions, directly implementing the pulses given in Eq. (14) cannot achieve high-fidelity gate operations. This is because an additional phase, independent of the geometric phase, accumulates on the states coupled to the external fields during the gate period. This unwanted phase depends purely on the frequency detuning.

To address this issue, we introduce an extra pair of pulses, referred to as compensation pulses, which swap the coupled and decoupled states and allow the system to evolve identically in both configurations. As a result, the originally decoupled state acquires the same detuning-dependent phase as the coupled one, turning it into a global phase and thereby recovering high fidelity.

This approach has been successfully validated in single-qubit gate systems [31]. Here, we extend the strategy to the two-qubit gate scenario by introducing two additional segments following Eqs. (8) and (9). The segments are configured with parameters $\theta' = \pi - \theta$, $\phi' = \phi + \pi$, $\gamma = 0$ and $\Omega_t(t)$ as follows:

$$\Omega_t(t) = \frac{\pi}{2\tau} + \sum_{k=1}^{s} a_k \frac{k\pi}{\tau} \cos\left(\frac{k\pi(t-2\tau)}{\tau}\right), t \in (2\tau, 4\tau), \quad (17)$$

where $a_k$ in Eq. (17) are the same as those in Eq. (14). These configurations effectively swap the states $|1b\rangle$ and $|1d\rangle$ during the additional evolution period, thereby enhancing the robustness of quantum manipulation where the control qubit is in state $|1\rangle$. In the case where the control qubit is in $|0\rangle$, both $|0b\rangle$ and $|0d\rangle$ remain coupled throughout the evolution governed by Eq. (12) and accumulate identical frequency-detuning dependent phases, making additional compensation unnecessary.

## D. Optimization of the Pulses

To enhance the gate performance respect to the frequency detuning and off-resonant excitation, we optimize the coefficients $a_k$ in Eqs. (14) by numerically solving the Lindblad-form master equation [38]:

$$\dot{\rho}(t) = -i[H'(t), \rho(t)] + \sum_{\alpha=1,2,3}\left\{L_\alpha \rho(t) L_\alpha^\dagger - \frac{1}{2}[L_\alpha^\dagger L_\alpha \rho(t) + \rho(t) L_\alpha^\dagger L_\alpha]\right\}, \quad (18)$$

where $\rho(t)$ denotes the density matrix of the system, and $H'(t) = H(t) + H_\Delta$ represents the total Hamiltonian. It includes both the resonant Hamiltonian $H(t)$ given in Eq. (6) in the basis of $\{|0\rangle, |1\rangle, |e\rangle\}$, and the detuning error term $H_\Delta = \Delta(|0e\rangle\langle 0e| + |1e\rangle\langle 1e| +$

$|e0\rangle\langle e0| + |e1\rangle\langle e1|)$. Details about $H_\Delta$ can be found in Supplement 1. In Eq. (18), $L_\alpha$ donates the jump operator. For the REI system, these operators are defined as follows:

$$L_1 = \sum_{l=c,t} \sqrt{\Gamma_1}(|0\rangle_l\langle e| + |1\rangle_l\langle e|), \tag{20}$$

$$L_2 = \sum_{l=c,t} \sqrt{\Gamma_2}(|e\rangle_l\langle e| - |0\rangle_l\langle 0| - |1\rangle_l\langle 1|), \tag{21}$$

and

$$L_3 = \sum_{l=c,t} \sqrt{\Gamma_3}|0\rangle_l\langle 1|, \tag{22}$$

where $l = c, t$ donates the control and target qubit. Here $\Gamma_1$ and $\Gamma_2$ represent the decay and pure dephasing rates from the excited state, respectively, while $\Gamma_3$ denotes the relaxation between the qubit levels. Here $\Gamma_3$ is set to zero, as the coherence time of $|0\rangle \leftrightarrow |1\rangle$ transition is significantly longer than those of the $|0\rangle \leftrightarrow |e\rangle$ and $|1\rangle \leftrightarrow |e\rangle$ transitions[39-41].

The optimal values of coefficients $a_k$ are determined using a multi-objective genetic algorithm (GA) framework based on Ref. [23]. This optimization approach simultaneously addresses two competing objectives: maintaining gate fidelity close to unity within a specified frequency detuning range ($|\Delta| \leq 170$ kHz in this work), and minimizing off-resonant excitations in ions far-detuned from the target transition ($|\Delta| \geq 8.9$ MHz). The solution yields a Pareto front, from which the final set of $a_k$ values is selected.

## 3. SIMULATION RESULTS

All simulations were performed by solving the Lindblad master equation (Eq. (18)) using the ode45 solver in MATLAB. Here the ensemble qubit in a $Eu^{3+}:Y_2SiO_5$ crystal is used as an example to demonstrate the performance of the gates, where the longitudinal and transverse relaxation rates are $\Gamma_1 = 2\pi \times 80$ Hz and $\Gamma_2 = 2\pi \times 60$ Hz[39], respectively. The evolution time is set to $\tau = 0.5$ μs, and the maximum instantaneous amplitude of Rabi frequency does not exceed 3 MHz. Given the strong dependence of the gate performance on the control qubit state ($|0\rangle$ or $|1\rangle$), a hybrid objective function was adopted in the optimization—combining 60% of the objective evaluated with input $|0\rangle$ and 40% with input $|1\rangle$—to obtain a balanced overall solution. The optimized values of $a_k$ are listed in Table 2, where $a_1 \sim a_4$ and $a_5 \sim a_8$ represent the coefficients for the two segments defined in Eqs. (8) and (9), respectively.

Table 2. Optimized values of PCH pulses for the CNOT and CS gates

|      | $a_1$ | $a_2$  | $a_3$  | $a_4$  | $a_5$  | $a_6$  | $a_7$ | $a_8$  |
|------|-------|--------|--------|--------|--------|--------|-------|--------|
| CNOT | 0.201 | −0.412 | −0.067 | −0.081 | −0.192 | −0.230 | 0.064 | −0.010 |
| CS   | 0.014 | −0.361 | −0.005 | 0.055  | −0.002 | −0.410 | 0.001 | 0.080  |

The temporal evolution of the Rabi frequencies, obtained using the optimized parameters $a_k$, is presented in Fig. 3, where (a) is for a CNOT gate, and (b) is for a CS gate. The solid-green, dashed-blue and dotted-red curves correspond to $\Omega_c(t)$, $\Omega_{0t}(t)$ and $\Omega_{1t}(t)$, respectively. The vertical dashed-blue line indicates the boundary between the gate-implementation and compensation pulses. All pulses evolve smoothly in time and vanish at their endpoints.

The performance of a CNOT and CS gate implemented with the pulses shown in Fig. 3 is evaluated below. A dipole-dipole interaction strength of $V = 50$ MHz is used, which is approximately 16 times the maximum instantaneous Rabi frequency.

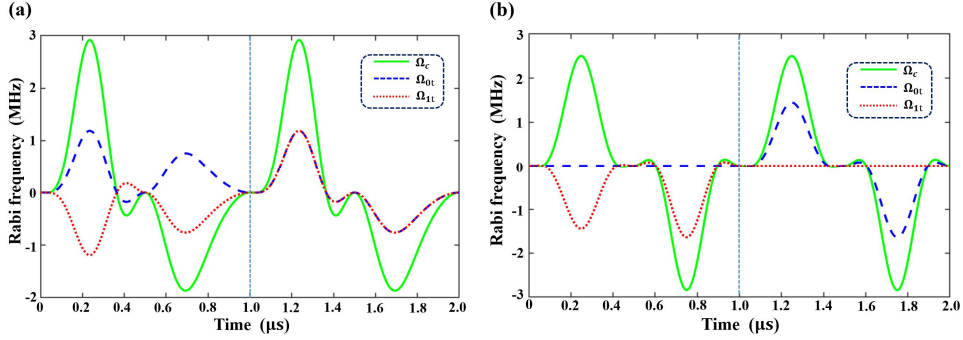

**Fig. 3.** Temporal evolution of the Rabi Frequency for the optimized PCH pulses corresponding to (a) a CNOT gate and (b) a CS gate. The solid green line, dashed blue line, and the dotted red line represent the control field $\Omega_c(t)$, the target fields $\Omega_{0t}(t)$, and $\Omega_{1t}(t)$, respectively. The vertical dashed blue line indicates the transition between the gate-implementation and compensation pulses.

## A. Gate operational fidelity

The performance of CNOT and CZ gates under ideal resonance conditions ($\Delta = 0$ in Eq. 19) is summarized in a truth table shown in Fig. 4. The initial state is either one of the four computational basis states ($|00\rangle$, $|01\rangle$, $|10\rangle$, or $|11\rangle$). In all cases, the gate fidelities exceed 99%. For the $|00\rangle$ and $|01\rangle$ inputs, the fidelity is limited by the strength of the dipole–dipole interaction, and is slightly lower than that for the $|10\rangle$ and $|11\rangle$ inputs, because the dipole-dipole interaction is not activated in the latter case. Implementing the above gate operation scheme in Rydberg-atom systems, where the interaction strength $V$ can reach several hundred MHz, would further increase the fidelities to about 99.8%, in close agreement with the values reported in Ref. [24]. It should be noted, however, that in the referred scheme only specific controlled two-qubit gates, such as CNOT, CH and CZ, can be realized.

Beyond computational basis states, the proposed scheme also performs effectively for arbitrary superposition input states. To evaluate this, we tested 1000 randomly generated superposition states of the form $\psi_s = a|00\rangle + b|01\rangle + c|10\rangle + d|11\rangle$, with $|a|^2 + |b|^2 + |c|^2 + |d|^2 = 1$, for a CNOT and a CS gate. The average gate fidelities obtained were 99.47% and 99.42%, respectively.

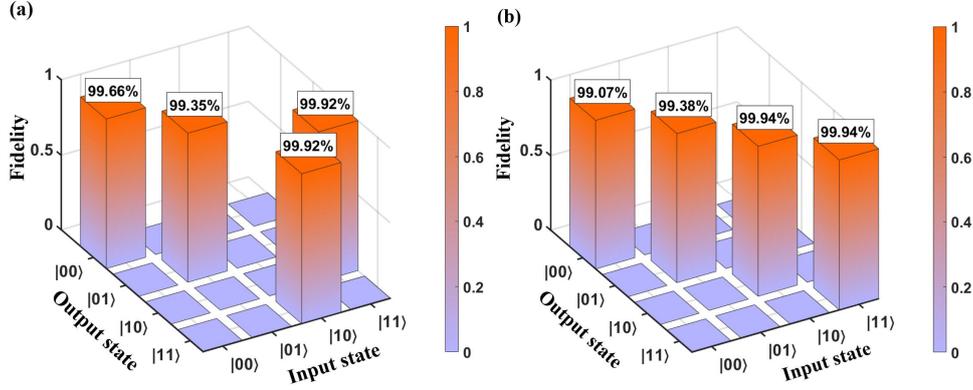

**Fig. 4.** Simulated truth tables for the two-qubit for the (a) CNOT gate and (b) a CS gate. The results, obtained with the optimized PCH pulses, show the fidelity of the output state in the computational basis ($|00\rangle$, $|01\rangle$, $|10\rangle$, $|11\rangle$) for each input.

The simulation results shown above confirm that the proposed scheme can indeed implement high-fidelity two-qubit gate operations in an ideal resonant system.

### B. High Robustness and low off-resonant excitation

In practical experimental environments such as ensemble qubits in a $Eu^{3+}:Y_2SiO_5$ crystal, ideal resonance conditions are often unattainable. High-fidelity gate operations in such systems therefore impose two essential requirements: the gate must exhibit strong robustness against frequency detuning within a range of ±170 kHz around resonance, while simultaneously maintaining sufficiently low off-resonant excitations for ions with transition frequencies detuned beyond 8.9 MHz from the target. To assess the applicability of our scheme in such systems, we evaluated both the robustness and off-resonant excitation characteristics for both the CNOT and CS gates.

The dependence of the gate fidelity on frequency detuning is presented in Fig. 5 for the CNOT gate (panels a and b) and the CS gate (panels c and d). Here (a) and (c) correspond to the initial state $|01\rangle$, while (b) and (d) correspond to the initial state $|11\rangle$. The solid red curves show the simulated fidelity obtained using the optimized coefficients $a_k$ listed in Table 2. The average fidelities within the ±170 kHz detuning range reach 99.24% (99.16%) for the CNOT(CS) gate with initial state $|01\rangle$, and 99.68% (99.64%) with initial state $|11\rangle$. For comparison, the dash-dotted blue curves depict results using a set of random parameters that maintain nearly the same maximum instantaneous Rabi frequency (all $a_k = 0$ except $a_1 = 0.45$, $a_3 = -0.15$, $a_4 = a_8 = -0.125$). Under the same detuning conditions and initial state $|01\rangle$, the average fidelity drops to 98.29% (98.77%) for the CNOT(CS) gate, approximately 0.5% lower than the optimized case. For initial state $|11\rangle$, the corresponding fidelities are 99.02% (CNOT) and 98.72% (CS), showing a reduction of about 0.7% compared to the optimized case. We should note that the slightly lower fidelity for $|01\rangle$ originates from the weakened rotating wave approximation under the current interaction strength. This influence will be analyzed in detail in Section D. In summary, parameter optimization consistently enhances robustness across different initial states, which is a key feature for implementing reliable quantum gates in realistic experimental environments.

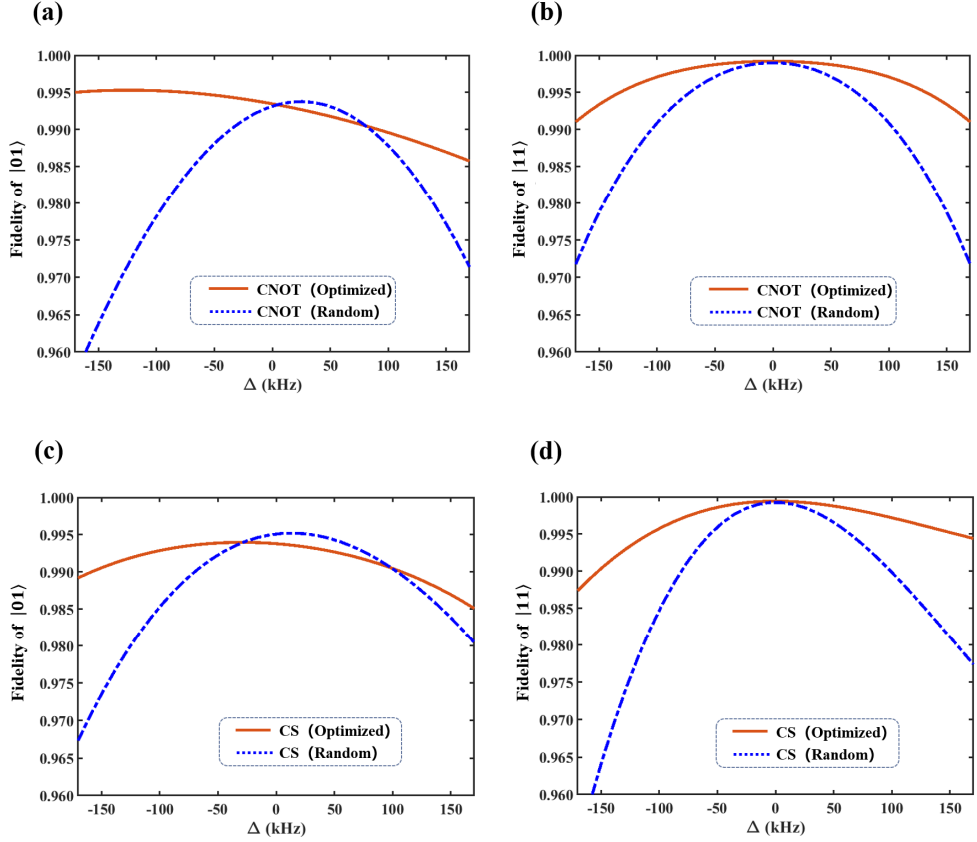

**Fig. 5.** Dependence of the operational fidelity on frequency detuning $\Delta$ for the CNOT gate (a and b) and CS gate (c and d). The fidelity is evaluated for two input states: $|01\rangle$ (a and c) and $|11\rangle$ (b and d). Results obtained with the optimized and unoptimized PCH pulses are represented by solid red and dashed blue curves, respectively.

The off-resonant excitation, defined as the total population transferred from the initial state to unintended levels, is plotted in Fig. 6 over a detuning range of 8.9-10 MHz. The panel layout and curve styles follow exactly the same as those in Fig. 5. The maximum off-resonant excitation beyond 8.9 MHz is 0.022% (0.035%) for the CNOT(CZ) gate with initial state $|01\rangle$, and 0.013% (0.004%) with initial state $|11\rangle$. In comparison, using the random set results in the off-resonant excitation roughly one order of magnitude higher, regardless of the specific gates and input states. These findings confirm that the parameter optimization effectively suppresses unwanted off-resonant excitation.

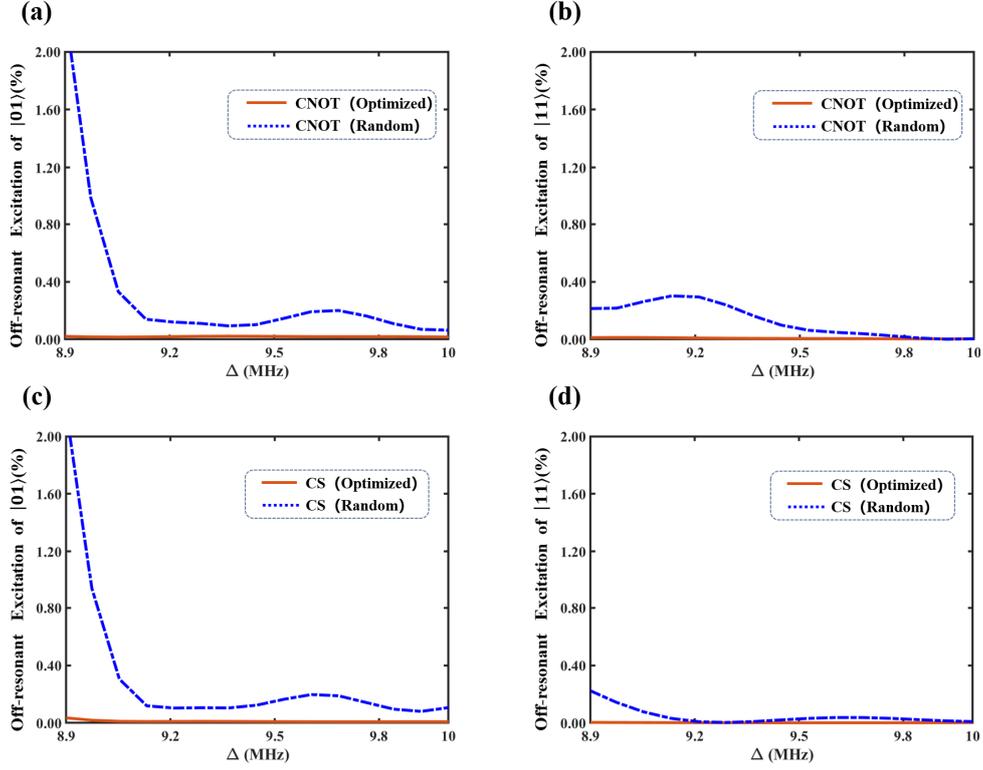

**Fig. 6.** Off-resonant excitation for the CNOT (a and b) and CS gates (c and d) in the frequency detuning range of [8.5, 10] MHz. The off-resonant excitation is evaluated for two input states: $|01\rangle$ (a and c) and $|11\rangle$ (b and d). Results obtained with the optimized and unoptimized PCH pulses are represented by solid red and dash-dotted blue curves, respectively.

## C. Performance of other gates

In addition to the CNOT and CS gates discussed above, we also investigated the performance of CH, CZ, and CT gates. Using the same optimization procedure, the resulting optimal pulse parameters are listed in Table 3, and the corresponding simulation results are summarized in Table 4. All gates achieve average fidelities above 99% within a detuning range of ±170 kHz, while the maximum off-resonant excitation remains well below 1%. These results demonstrate that the proposed approach is effective for implementing arbitrary controlled two-qubit gates in the ensemble $Eu^{3+}:Y_2SiO_5$ qubit system.

**Table 3. Optimized parameters in the PCH pulses for other controlled gates**

|    | $a_1$  | $a_2$  | $a_3$  | $a_4$ | $a_5$  | $a_6$  | $a_7$  | $a_8$  |
|----|--------|--------|--------|-------|--------|--------|--------|--------|
| CH | 0.110  | −0.417 | −0.037 | 0.083 | −0.408 | −0.138 | 0.136  | −0.056 |
| CZ | 0.259  | −0.482 | −0.086 | 0.116 | −0.438 | −0.187 | 0.146  | −0.031 |
| CT | −0.294 | −0.293 | 0.098  | 0.022 | 0.452  | −0.395 | −0.151 | 0.072  |

**Table 4. The average operational fidelity and off-resonant excitation of gates CH, CZ, and CT for input states $|01\rangle$ and $|11\rangle$ using the optimal pulses**

|    | $\bar{F}_{|01\rangle}$ | $\bar{F}_{|11\rangle}$ | $P_{\max\_|01\rangle}$ | $P_{\max\_|11\rangle}$ |
|----|------|------|------|------|
| CH | 99.45% | 99.87% | 0.142% | 0.072% |
| CZ | 99.52% | 99.85% | 0.185% | 0.049% |
| CT | 99.22% | 99.63% | 0.084% | 0.007% |

### D. Impact of the strength of the dipole-dipole interaction

All simulation results presented above are obtained with a dipole–dipole interaction strength of $V = 50$ MHz, approximately 16 times the maximum Rabi frequency. As expected, higher gate fidelities can be achieved with larger values of $V$, owing to a more pronounced dipole-blockade effect. To quantify this dependence, we evaluated the fidelities of the CNOT and CS gates for the input state $|01\rangle$ across a range of interaction strengths; the resulting fidelity–interaction relation is shown in Fig. 7. Panel (a) corresponds to the CNOT gate, and panel (b) to the CS gate.

As shown in Fig. 7, the gate fidelity increases steadily with $V$, although the rate of improvement gradually saturates with increasing interaction strength. In practice, however, $V$ is often constrained by experimental conditions. For example, in ensemble REI systems, $V$ typically remains below 100 MHz due to limitations in material fabrication. The value of 50 MHz adopted here is consistent with theoretically estimates reported in Ref. [42]. It is worth noting that the upper limit of $V$ in the ensemble REI systems could be raised in future through techniques such as ion implantation for highly doped regions [43] or optical cavity enhancement [44]. Such advances would enable higher performance while remaining compatible with the proposed scheme.

Furthermore, the trend observed in Fig. 7 underscores the scalability of our scheme to platforms with intrinsically stronger interactions. In Rydberg-atom architectures, for instance, where $V$ routinely reaches hundreds of megahertz, our protocol is expected to deliver even higher fidelities due to the enhanced blockade effect. This demonstrates the generality of the proposed method and its adaptability to diverse physical platforms.

In summary, the proposed two-qubit gate scheme achieves high-fidelity performance with low-power driving, while exhibiting high robustness against frequency detuning and low off-resonant excitation of neighboring qubits. These results demonstrate that the protocol provides an experimentally feasible pathway to realize arbitrary controlled two-qubit gates in systems such as REIs, offering both high precision and intrinsic robustness essential for scalable quantum computing.

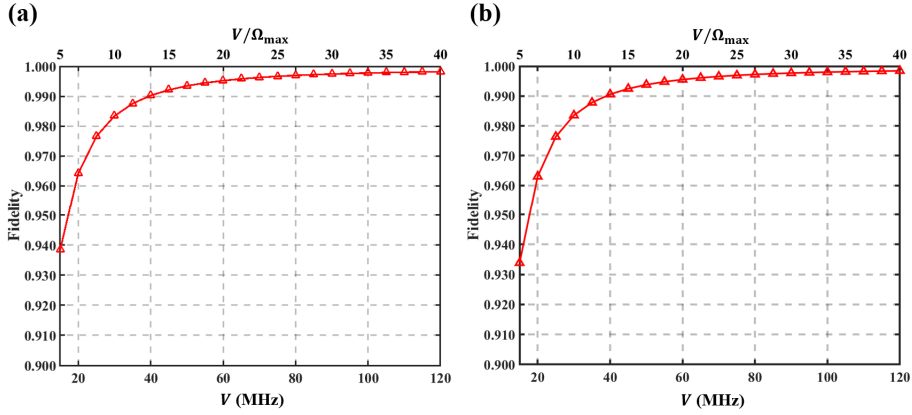

**Fig. 7.** Dependence of the gate fidelity on the strength of the dipole–dipole interaction $V$ for the (a) CNOT gate and (b) CS gate, evaluated for the input state $|01\rangle$. The top axis denotes the normalized strength $V/\Omega_{\max}$.

## 4. CONCLUSION

We propose a resonant scheme for high-fidelity arbitrary controlled two-qubit gates in experimental platforms dominated by dipole-dipole interactions (e.g., rare-earth-ion-doped crystals). A key feature of this scheme lies in the adoption of asymmetric excitation, which constructs independent evolution pathways for the control and target qubits. This design enables their simultaneous yet decoupled manipulation through a two-segment pulse sequence. The resonant nature of the approach alleviates the stringent requirement for precise detuning control and shortens the gate operation time. By integrating optimized PCH pulses with compensation pulses, the scheme maintains a fidelity exceeding 99% within a range of ±170 kHz, while suppressing off-resonant excitation to below 0.2%. The proposed scheme is also applicable to other dipole-interaction–dominated systems, such as the Rydberg atoms by tailoring the pulse optimization objectives to prioritize the specific requirements. This work validates the feasibility of a resonant route for non-time-sequenced two-qubit gates, breaking the reliance on off-resonant control in such gate architectures. Furthermore, it underscores the compatibility and potential of pulse engineering in rare-earth-ion systems, laying a foundation for future high-fidelity quantum gate implementations on this platform.

**Funding.** This work was supported by the China Scholarship Council (CSC), the Natural Science Foundation of Jiangsu Province (No. BG2025017 and No. BK20250404), the Youth Science and Technology Talent Support Project of Jiangsu Province, and the Frontier Technology Research Program of Suzhou (No. SYG202322).

**Disclosures.** The authors declare no conflicts of interest.

**Data availability.** The datasets utilized and analyzed during the current study are available from the corresponding author upon reasonable request.

**Supplemental document.** See Supplement 1 for supporting content.